# Advancing Health Equity through Multi-Level Fairness in Health Informatics

Nick Souligne*

College of Engineering, University of Arizona, Tucson, AZ, nasouligne@arizona.edu

Vignesh Subbian

College of Engineering, University of Arizona, Tucson, AZ, vsubbian@arizona.edu

The increasing integration of machine learning in healthcare has highlighted critical challenges related to fairness, transparency, and health equity. Specifically, the use of multi-level fairness techniques, which combine multiple bias mitigation steps or techniques, show promise for reducing biases across different patient demographics, yet this approach remains underexplored in terms of its health equity outcomes. In this paper, we assess the current landscape of multi-level fairness in health informatics by focusing on its impact on equitable healthcare outcomes and evaluating how transparency and reporting standards contribute to these advancements. Through an examination of the existing literature, we identify key gaps in both the implementation of multi-level fairness techniques and the consistent reporting of health equity impacts. Furthermore, we analyze the role of reporting standards, including MINIMAR and TRIPOD, in improving model transparency and ensuring that machine learning models in healthcare address health disparities. These standards offer valuable benchmarks for reporting on ML models, yet we identify key opportunities for enhancing how these reports capture fairness and equity outcomes. The paper concludes by providing recommendations that focus on improving transparency in reporting, advocating for the broader adoption of multi-level fairness techniques, and ensuring that health equity is explicitly prioritized in future research efforts.

## 1 Introduction

Machine learning (ML) and artificial intelligence (AI) are increasingly shaping critical aspects of society, with profound implications on healthcare, legal, and social outcomes. In healthcare, the use of ML is often hailed for its potential to enhance clinical decision-making and streamline care. However, these technologies frequently rely on real-world data, such as Electronic Health Records (EHRs), which can reflect and perpetuate societal biases. When applied in clinical settings, these biases can lead to unequal treatment recommendations, misdiagnoses, or worse outcomes for historically underserved and underrepresented populations. As a result, ML-driven healthcare risks exacerbating existing health inequities, raising urgent questions about how to ensure these technologies promote equitable, rather than harmful, health outcomes.

Biases in ML models arise in several forms, often reflecting or amplifying existing health inequities. There are many different types of bias, but the most relevant in biomedical applications tend to be historic, representation, aggregation, population, or measurement bias. Historic bias stems from societal inequities embedded in the data used for training, while representation bias occurs when data underrepresents marginalized groups, skewing outcomes. Aggregation bias results from oversimplifying assumptions applied to individuals or subgroups, and population bias arises when the dataset's patient population differs from the target population, reducing model generalizability. Lastly, measurement bias occurs when features used in the model are not equally relevant for all groups, further distorting predictions [1]. These biases can undermine the fairness of ML systems, making techniques for detecting and mitigating them, often referred to as Algorithmic Fairness, essential. Addressing these biases is critical to preventing ML from perpetuating disparities in healthcare.

Bias mitigation strategies in ML typically fall into three categories: pre-processing, in-processing, and post-processing. Pre-processing focuses on cleaning and adjusting the data before it is fed into the model, aiming to prevent the model from inheriting or learning harmful biases. In-processing involves mitigating bias during the model's training phase, using techniques like fairness constraints or regularization terms to ensure more equitable outcomes. Post-processing seeks to correct bias after the model has been trained, adjusting predictions or outcomes to enhance fairness.

While these techniques offer a starting point, bias mitigation in clinical informatics must be approached comprehensively. Prior reviews have examined the impact of bias in ML models and evaluated various mitigation strategies, particularly for models trained on EHR data [6]. However, these reviews often overlook the importance of addressing bias across multiple levels of the development process. As highlighted by Rana, Azizul, and Awan, a heterogenous approach to clinical ML is required, targeting bias in the data, algorithms, and decision-making processes, and ensuring that development cycles are aligned [3]. A multi-level approach to fairness—one that integrates mitigation strategies throughout the ML lifecycle and is reinforced by robust reporting standards—offers a more effective path toward advancing health equity. To be successful, this approach must be multifaceted, adaptable, transparent, scalable, accessible, and interdisciplinary, with iterative refinement as new evidence emerges [3].

*Corresponding Author

There remain significant gaps in the current literature on multi-level fairness in ML models, especially within the context of health informatics. Our goal is to identify these gaps and propose actionable recommendations for future research. Specifically, we aim to examine health equity through the lens of ML models, assess the current literature on multi-level fairness, and suggest ways to improve reporting on fairness. Ultimately, we will advocate for future ML models that are designed with health equity as a central consideration, ensuring that technology becomes a tool for closing, rather than widening, the health disparity gap.

## 2 Background

The growing interest in machine learning (ML) systems, particularly in health informatics, has sparked numerous studies on bias mitigation strategies. Unfortunately, many of these studies address bias mitigation without fully considering the actual impacts on health equity. Ensuring fairness in ML models, particularly those used in healthcare, is not just a technical challenge, it is an ethical imperative. When models are trained on biased data, they risk amplifying existing disparities, disproportionately impacting marginalized populations and reinforcing unequal healthcare outcomes. Thus, addressing bias in ML is essential to promoting justice and equity in healthcare.

The pursuit of health equity in ML-driven healthcare systems demands a multi-level approach. As highlighted in previous work, bias mitigation strategies can be classified into three broad categories: pre-processing, in-processing, and post-processing. Hort et al. [2] provided an in-depth overview of these categories, detailing how each strategy functions within the model development cycle. Furthermore, Hort et al. reviewed 341 studies, of which 123 utilized a pre-processing approach, 212 employed in-processing techniques, and 56 relied on post-processing methods. Additionally, 70 studies used a combination of two or more techniques, including only 4 studies that combined pre-processing and post-processing. These findings highlight gaps in the literature, particularly the limited exploration of comprehensive, multi-level strategies.

Despite the volume of research, very few studies directly consider the implications of their findings for health equity. Studies such as Wan et al. [4] and Caton and Haas [5] provide comprehensive reviews of processing techniques and fairness in ML but often lack a specific focus on how these techniques impact underserved populations. Similarly, while others such as Huang et al. [9] and Drabiak et al. [10] delve into the application of real-world data and the ethical challenges of ML, they rarely address whether these models meaningfully improve or exacerbate health disparities. As a result, while these reviews offer valuable insights into fairness metrics and technical approaches, they stop short of fully integrating health equity into the conversation.

Though some studies have made strides in connecting bias to health equity outcomes, much work remains to explicitly bridge this gap. For example, Young et al. [8] examined the relationship between measurement error and bias in healthcare models, while Gichoya et al. [11] highlighted biases as a common pitfall of AI in clinical settings. However, comprehensive frameworks that target health equity throughout the entire ML development process are still needed as ML becomes more integral to healthcare decision-making.

A key area where fairness and health equity concerns converge is in how ML models are designed and reported. Transparency in reporting, particularly regarding how models are developed, tested, and validated, plays a crucial role in ensuring that bias mitigation strategies are effective. Reporting must be more than a description of technical performance; it must include details on how health equity considerations are woven into the model development cycle. For instance, which populations are represented in the training data? Are marginalized groups adequately included, or are they being systematically underrepresented? How are fairness metrics selected and measured, and do they reflect the realities of those most affected by healthcare disparities?

Guidelines for promoting fairness in ML, such as the Fairness of Artificial Intelligence Recommendations (FAIR) framework [7], provide valuable starting points. However, many of these guidelines neglect the importance of addressing bias at multiple stages of model development. Relying on interventions at a single stage is insufficient for achieving health equity. Instead, comprehensive frameworks that combine pre-processing, in-processing, and post-processing techniques, along with a commitment to transparent and accountable reporting, are essential for reducing disparities in health outcomes and advancing equity in healthcare.

## 3 Focus 1: Health Equity

Health equity refers to the principle of providing fair and just opportunities for all individuals to achieve their highest potential for health, regardless of social, economic, or demographic factors [18]. It acknowledges that certain populations such as racial and ethnic minorities, low-income individuals, and those in rural or underserved areas often face systemic barriers that prevent them from accessing high-quality healthcare. These inequities manifest in various forms, from disparities in access to care and medical resources to unequal treatment and outcomes in healthcare settings.

### 3.1 Who Does It Affect and Why?

Health inequities disproportionately affect historically marginalized groups. For instance, Black, Indigenous, and People of Color (BIPOC) often experience worse health outcomes due to the legacy of structural racism, which has historically shaped everything from the quality of healthcare facilities to the biases embedded in diagnostic tools [16,17]. Low-income populations also face challenges such as underinsurance or lack of access to preventative care, which can lead to a higher burden of chronic disease. These disparities are further exacerbated by social determinants of health, including education, employment, and housing conditions, which are often shaped by policies and practices that have historically disadvantaged these groups [18].

### 3.2 How Can Health Equity Be Assessed?

Assessing health equity requires a comprehensive evaluation of both healthcare processes and outcomes across different population groups, with an intersectional approach being particularly crucial. Health inequities often emerge not just along a single axis such as race or socioeconomic status, but through the interplay of multiple social identities, such as gender, disability, and ethnicity. Intersectionality helps reveal how overlapping social categories can compound disadvantages due to biases, leading to unique barriers to healthcare access and outcomes. Therefore, identifying and addressing these compounded inequities is key to a meaningful assessment of health equity.

Traditional metrics such as access to services, quality of care, patient satisfaction, and clinical outcomes such as differences in mortality rates or disease prevalence are foundational for identifying disparities. However, in the context of machine learning models, fairness metrics provide additional insights into whether these tools are promoting or undermining equity. Fairness metrics often focus on ensuring that model performance is consistent across different demographic groups. Commonly used metrics include demographic parity, which measures if a model's positive prediction rate is equal across groups, and equalized odds, which ensures parity in both false positive and false negative rates across groups. Predictive value parity, which measures whether the accuracy of predictions is the same for different groups, is another crucial metric for assessing fairness.

### 3.3 Linking Fairness Metrics to Health Equity Outcomes

While fairness metrics can help ensure that models function without obvious biases, they alone are insufficient for achieving true health equity. To fully link fairness metrics to health equity outcomes, it is important that the fairness of model predictions translates into equitable access to care, quality of treatment, and healthcare outcomes for all groups, including those impacted by multiple, intersecting forms of bias. For instance, a model might achieve demographic parity by producing similar prediction rates across racial groups, but if it fails to account for the compounded disadvantages faced by, say, low-income Black women, it may still fail to advance equity in a meaningful way. Similarly, even if a predictive model identifies disease risk equally across racial groups, if structural inequalities in the system where the model is deployed still prevent BIPOC patients from accessing high-quality treatment, the model's fairness metrics will have little impact on actual health equity. Therefore, assessing health equity in machine learning models requires both technical fairness and an intersectional evaluation of how those predictions are acted upon within healthcare systems. This includes tracking whether models lead to equitable follow-up care, resource allocation, and outcomes for people facing overlapping social barriers.

A major challenge in achieving health equity is that fairness interventions must account for multiple levels of bias, from individual-level discrimination to systemic inequities embedded in healthcare infrastructures. Addressing these biases requires multi-level fairness frameworks, which extend beyond single-axis fairness metrics to assess and mitigate disparities at different structural levels including clinical decision-making, and policy implementation. Adopting multi-level fairness enables researchers and practitioners to develop machine learning models that are both technically fair and contextually aligned with the systemic inequities they aim to address. As illustrated in Figure 1, health equity serves as the driving force behind the need for multi-level fairness frameworks. These frameworks, in turn, depend on transparency and reporting standards to evaluate and refine fairness interventions, ultimately fostering greater realization of health equity.

However, ensuring multi-level fairness requires clear transparency and reporting standards that allow researchers, clinicians, and policymakers to evaluate whether fairness interventions are effective. Without robust reporting mechanisms, fairness assessments risk being ad hoc, inconsistently applied, or disconnected from actual health equity outcomes. Transparent reporting frameworks help ensure that fairness evaluations are systematically documented, reproducible, and accountable to the communities they impact.

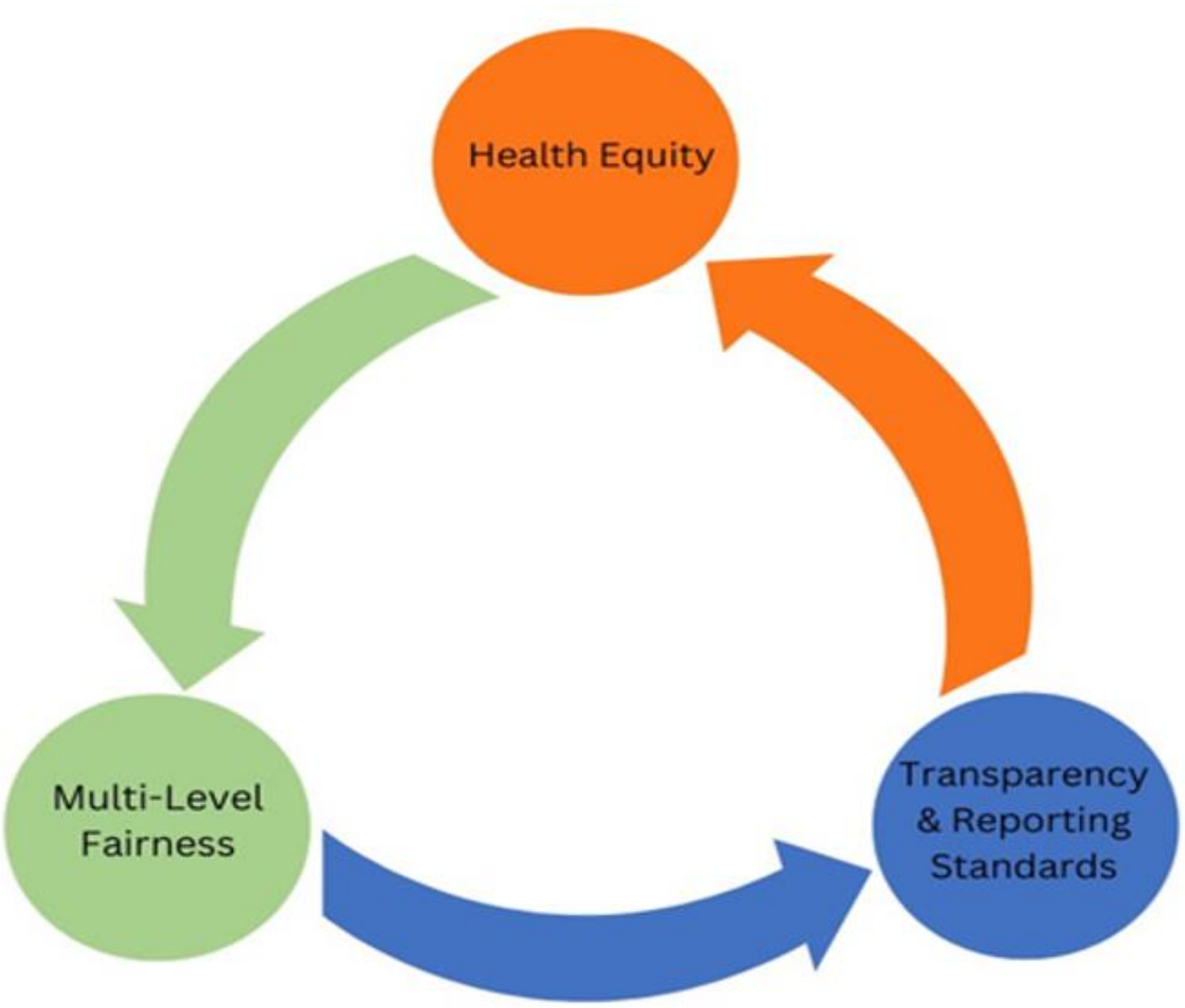

**Figure 1. Health Equity drives the need for Multi-Level Fairness, which in turn drives the need for Transparency and Reporting standards leading to greater realization of health equity**

Ultimately, ensuring health equity involves more than mitigating bias within the algorithm itself; it also requires that fairness metrics be tied to actionable, real-world improvements. Intersectional approaches to fairness ensure that machine learning models can recognize and address the compounded disadvantages faced by underserved groups, translating fairness into real-world health equity outcomes. This highlights the need to explore the various initiatives currently underway across different sectors that aim to promote health equity through targeted actions and innovative solutions.

### 3.4 Efforts to Improve Health Equity

Over recent years, there have been significant efforts to close the health equity gap across multiple sectors:

- **Government Initiatives:** Governments worldwide have introduced policies aimed at addressing disparities in healthcare. In the U.S., initiatives like *Healthy People 2030* have set specific goals for reducing healthcare disparities across various demographic groups and programs such as *All of Us* seek to create new repositories of clinical health data that better represent the country's diverse population. Similarly, the *Affordable Care Act* has expanded access to healthcare for millions of previously uninsured individuals, many of whom belong to marginalized communities [19-21].
- **Private Sector:** The private sector has also made strides toward improving health equity. Many healthcare organizations are investing in community health programs, expanding telemedicine services to reach underserved areas, and incorporating social determinants of health into patient care models. Moreover, pharmaceutical and technology companies are exploring how to use data-driven approaches, such as AI and ML, to identify and address health disparities. However, there is ongoing debate about whether these efforts are sufficient or if profit-driven motives sometimes undermine long-term equity goals [22,23].
- **Academic Efforts:** In academic circles, there has been a growing focus on developing fair and equitable machine learning models to support clinical decision-making. Studies, such as those by Wan et al. [4] and Caton and Haas [5], have laid important groundwork in understanding bias in ML, while efforts by Huang et al. [9] have begun to explore how fairness can be operationalized in real-world clinical settings. Still, much of the research has yet to fully integrate health equity as a core consideration, often focusing more on technical fairness metrics than on the tangible impacts these models have on marginalized groups. The literature, while expanding, is only beginning to grapple with the full complexity of health equity.

## 4 Focus 2: Multi-Level Fairness Techniques

In the pursuit of fairness in machine learning models, particularly within healthcare, multi-level fairness techniques have emerged as a promising approach to addressing biases at various stages of the model development process. Unlike single-level fairness approaches, which typically focus on correcting bias at a single point, multi-level fairness encompasses a more comprehensive strategy. It applies interventions at multiple stages, including pre-processing, in-processing, and post-processing, to reduce the risk of perpetuating or exacerbating health disparities across diverse populations.

Figure 2 illustrates that the effectiveness of fairness interventions varies based on the number and combination of mitigation strategies applied, which can be informally understood as the model fairness level. Models with no mitigation (Level 0) often retain high levels of bias, whereas those employing single (Level 1) fairness techniques, multi-level fairness techniques (Level 2), or multi-level fairness techniques informed by transparent reporting (Level 3) demonstrate increasingly equitable outcomes. This progression highlights the critical role of integrating multi-level fairness strategies to achieve meaningful reductions in disparities.

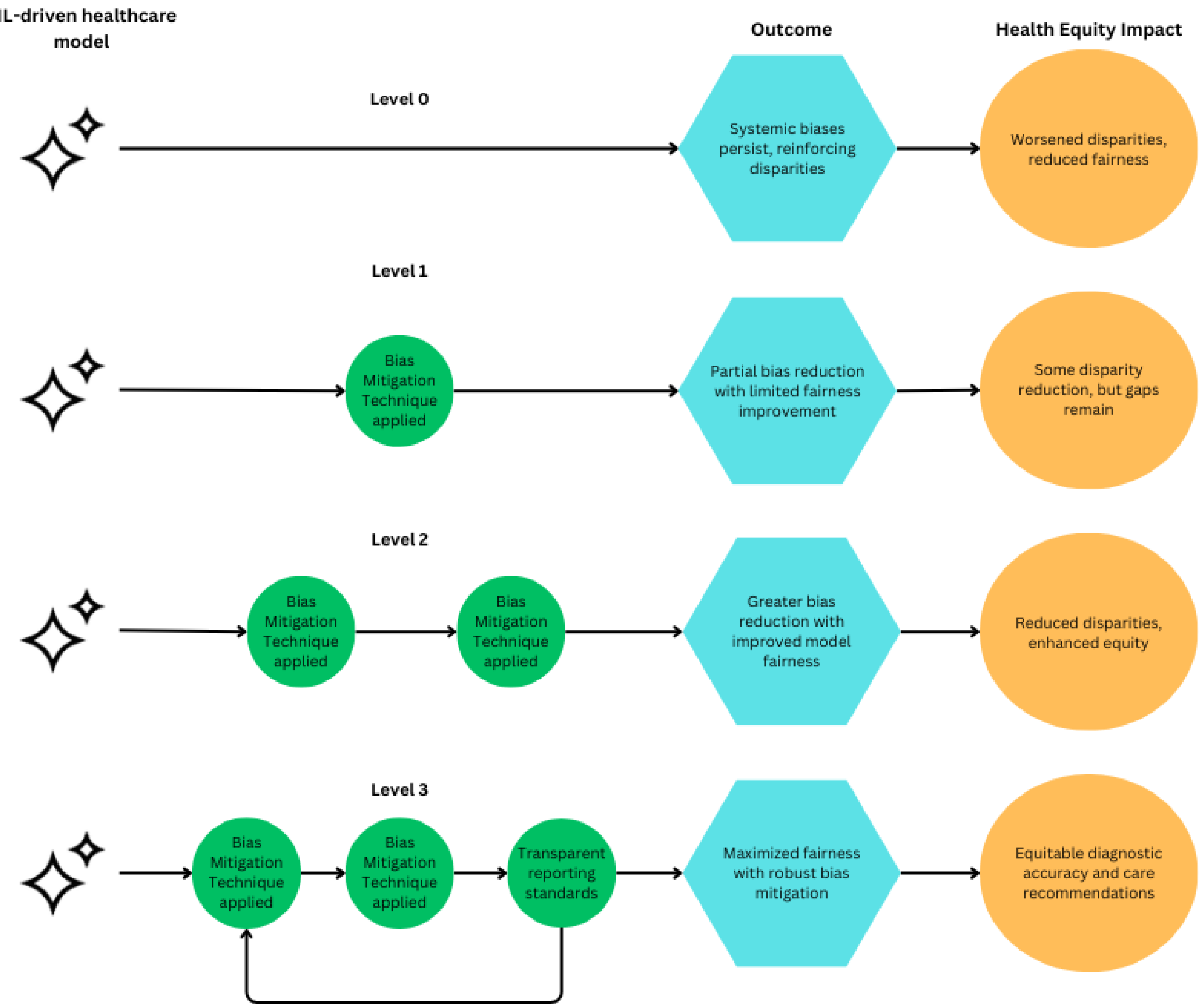

**Figure 2. Illustration how different levels of fairness interventions can affect bias reduction and overall health equity outcomes.**

### 4.1 Studies Utilizing Multi-Level Fairness Techniques

In reviewing relevant literature, we identified several studies that employed multi-level fairness techniques. While the specific contexts and models varied, a common theme was the combination of pre- and in-processing techniques. Notably, few studies, especially in the health informatics field, have investigated the effects of combining post-processing with other techniques [2]. Most of the literature applied fairness interventions across protected characteristics race, ethnicity, gender, and age. Another noteworthy commonality was the tension or trade-off between model accuracy and fairness. Unfortunately, there seemed to be no common standards followed when it came to reporting the features, characteristics, or outputs of the models or their development process.

### 4.2 Impacts of Multi-Level Fairness on Health Equity

Assessing the impacts of multi-level fairness techniques on health equity reveals a promising trend in addressing disparities in healthcare outcomes. For instance, one study focused on predicting post-liver transplant risk factors found that employing a multi-level approach with in- and post-processing techniques significantly enhanced equitable predictions across various subgroups, including age, gender, and race/ethnicity [13]. Other research similarly indicates that integrating multi-level fairness techniques can achieve substantial improvements in fairness between groups with only minimal trade-offs in model accuracy [14,15]. However, significant gaps persist in the literature. There is notable scarcity of comprehensive studies aggregating knowledge in this area, which may stem from the relative novelty of multi-level fairness approaches in health informatics. Despite this gap in the literature, extensive research has already been conducted on fairness in machine learning and health informatics more broadly. As researchers increasingly recognize the importance of integrating multiple levels of fairness techniques, the body of literature is likely to grow. Although many reviews encompass relevant studies, they often mix multi-level approaches with single fairness techniques and rarely concentrate specifically on health informatics. Furthermore, numerous studies on multi-level fairness do not adequately analyze how their advancements translate into meaningful improvements for the target populations. This highlights the need for standards for assessing and reporting these advancements.

## 5 Focus 3: Model Transparency and Reporting Standards

Evaluating the impact of multi-level fairness on health equity requires robust standards for reporting and model transparency. Transparent reporting plays a crucial role in ensuring that biases, limitations, and population-specific performance metrics are openly acknowledged and addressed, ultimately enhancing the interpretability and trustworthiness of ML models. Reporting standards guide researchers in systematically documenting key aspects of model development, evaluation, and deployment, providing a framework for accountability.

As shown in Figure 3, transparent reporting is not a standalone process but an integral part of an interconnected system that includes Bias Detection and Bias Mitigation. Together, these components form an implementation system where transparency reinforces bias detection by making disparities more visible and strengthens bias mitigation by enabling more effective and targeted interventions. This synergy ensures that fairness considerations are embedded throughout the ML lifecycle, promoting models that are both equitable and beneficial to all stakeholders [26].

Within this framework, we conceptualize additional roles as key stakeholders who contribute to the development and guidance of algorithmic fairness efforts. Researchers play a crucial role in developing and refining bias detection and mitigation strategies. Hospitals and clinicians serve as primary implementers, integrating fairness-driven practices into clinical workflows and decision-making processes. Patients provide essential feedback on how these practices impact their care, ensuring that fairness interventions align with real-world needs. Insurance companies shape fairness considerations through coverage policies and reimbursement structures. These roles interact closely with developers, who operationalize fairness measures within machine learning models, and with broader stakeholders, including policymakers and institutional leaders, who drive accountability and advocate for transparency.

Transparent reporting is crucial for enabling each of these stakeholders to effectively contribute to fairness efforts. For researchers, it ensures the rigorous evaluation of bias detection and mitigation strategies. For hospitals, clinicians, and insurance companies, it provides the clarity needed for informed decision-making and the integration of equity-focused practices. For patients, transparent reporting builds trust by making algorithmic decisions more understandable and accountable. By engaging all these actors, the implementation system ensures that fairness in machine learning is not merely a technical objective but a sustained and collaborative effort across the healthcare ecosystem.

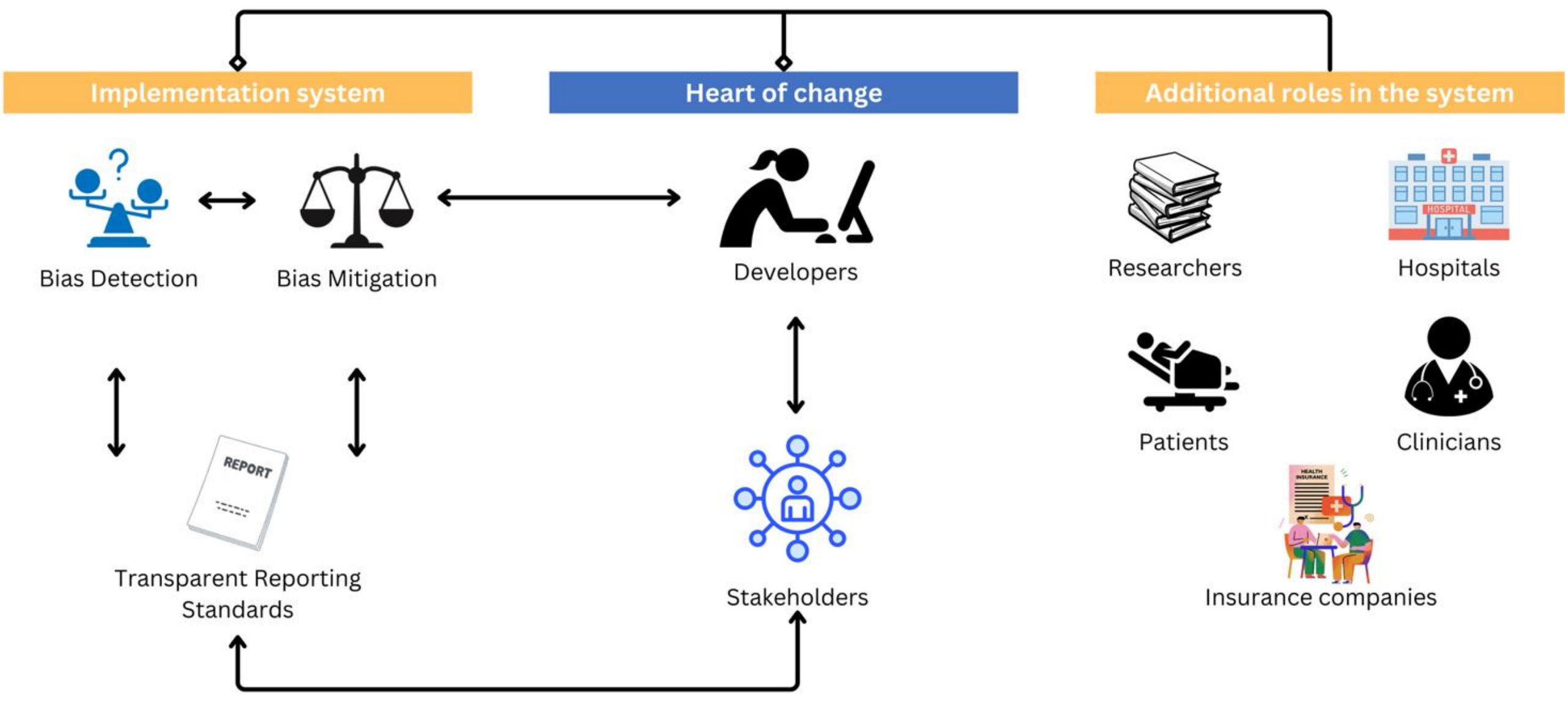

**Figure 3.** **Interactive Systems Framework showing how the interconnected systems (Bias Detection, Bias Mitigation, and Transparent Reporting Standards) work together to promote algorithmic fairness**

### 5.1 Current Reporting Standards

Numerous reporting standards have been established in the literature, as detailed in a systematic review by Kolbinger et al. [24]. Two key frameworks particularly relevant for multi-level fairness models are MINIMAR (Minimum Information for Medical AI Reporting) and TRIPOD-AI (Transparent Reporting of a Multivariable Prediction Model for Individual Prognosis or Diagnosis – Artificial Intelligence). Both standards provide a solid foundation for documenting key elements of model development and evaluation, requiring the reporting of study population, patient demographics, model architecture, and evaluation plans. Yet, important distinctions exist between the two. TRIPOD-AI, for instance, includes specific provisions for reporting class imbalance and fairness, which are critical for evaluating the equity of a model's performance. In contrast, MINIMAR focuses on the minimum reporting requirements and lacks detailed guidance on fairness, making TRIPOD-AI a more suitable choice for studies involving multi-level fairness. That said, while TRIPOD-AI offers a stronger framework for fairness reporting, it still has notable limitations. One significant gap is the absence of guidance on how to report post-deployment model evaluation, particularly in terms of their impact on equitable health outcomes. Furthermore, its requirements for fairness reporting are somewhat vague, failing to specify how steps to mitigate biases throughout the model development lifecycle should be documented. This leaves room for improvement in ensuring that fairness is adequately addressed and transparently reported across all phases of model development and use [12,25].

### 5.2 Impacts on Health Equity

Reporting standards play a crucial role in impacting health equity by ensuring that models are developed, evaluated, and deployed with transparency regarding their performance across diverse populations. The impact of these standards on efforts to improve health equity cannot be overstated, as they provide a structured framework for documenting essential details of the model-building process. This includes how data was collected, how demographic variables such as race, gender, and socioeconomic status are represented, and how models perform across different subgroups. Such transparency enables researchers, clinicians, and even policymakers to identify where biases may exist in model development and implementation, which is the first step toward mitigating those biases.

Through detailed reporting, standards like TRIPOD-AI help ensure that key fairness measures are documented, making it easier for future researchers to understand how biases arise in clinical AI models. For instance, models trained on datasets that underrepresent certain populations, such as racial minorities or low-income groups, may produce inequitable outcomes if these imbalances are not properly reported and addressed. Reporting standards encourage researchers to highlight these disparities, prompting more equitable data collection practices and model adjustments to avoid perpetuating existing health inequalities. Moreover, these standards ensure that efforts to improve fairness are not only made but also properly evaluated and shared with the broader scientific community. This transparency promotes continuous learning and improvement within the field of AI in healthcare, as researchers can build on each other's work, identifying best practices and innovative strategies for mitigating bias.

## 6 Recommendation 1: Emphasize Fairness throughout the model development lifecycle

The model development life cycle (MDLC) includes all stages of creating machine learning models, from data collection and preprocessing to training, evaluation, and deployment. Prioritizing fairness at each step is essential for improving health equity, as biases can emerge at any stage, leading to unequal model performance across diverse populations. Integrating fairness considerations throughout the MDLC helps ensure that models are more equitable and suitable for clinical applications impacting a wide range of patients.

During data collection, fairness can be promoted by aligning datasets with the characteristics of the target population where the model will be deployed, ensuring that diverse demographic groups such as age, race, gender, and socioeconomic status are appropriately represented. In preprocessing, imbalances from data collection should be addressed, whether through rebalancing techniques or other strategies to mitigate skewed representation. For model training, fairness-aware algorithms can detect and reduce biases, particularly those that may disproportionately affect specific groups. Special attention should be given to intersectional groups with overlapping protected characteristics to prevent disparate impact. In the evaluation phase, models should be assessed for their performance across all subgroups, with an emphasis on identifying and addressing any disproportionate effects on these intersectional groups. Deployment should include ongoing monitoring of model performance, with mechanisms in place to adjust the model if new biases or inequities arise over time.

Embedding fairness throughout the MDLC increases the likelihood that researchers and practitioners will create models that support health equity and reduce the risk of reinforcing existing disparities in clinical outcomes.

## 7 Recommendation 2: Expand Fairness Reporting

Currently, there are reporting standards in place for clinical models; however, these standards often fail to capture all the necessary nuances of fairness. Unfortunately, their adoption is limited as much of the literature does not utilize these standards, resulting in significant inconsistencies in the information reported. This lack of cohesion poses a serious challenge for the field, as it hampers the reproducibility of new methodologies and restricts findings to specific populations or settings. Our recommendation is two-fold: first, the field must reach a consensus on the reporting standards to be used; and second, this standard should include or expand to encompass specific items related to fairness and health equity.

We recognize that there are many different reporting standards in literature, and that many of them are specialized for a specific context. While it may not be possible to find a standard that fully encompasses every possible study in multi-level fairness for Health Informatics, that should not hinder the adoption of a general-purpose standard. In this paper, we have previously discussed TRIPOD-AI, and we recommend that it be considered the primary standard for multi-level fairness research. TRIPOD-AI encompasses many of the features that are most reported in the literature and includes several critical items, such as class imbalances and fairness measures, which are essential in health informatics. It is our hope that researchers will adopt this standard leading to greater cohesion in both multi-level fairness and the broader field of health informatics.

While TRIPOD-AI serves as a solid foundation for a reporting standard, additional elements should be included when reporting studies on multi-level fairness. Specifically, we recommend adding items for bias detection and mitigation strategies employed during data collection, model training, deployment, and evaluation. Furthermore, reporting should include steps taken to assess the impacts on health equity, particularly after model deployment. Lastly, guidance on how to modify or adjust the model post-deployment should also be incorporated.

## 8 Conclusion

In conclusion, the importance of fairness in health informatics cannot be overstated, especially as machine learning models become integral to healthcare decision-making. Significant gaps persist in the literature regarding fairness reporting, with many studies lacking standardized guidelines and comprehensive assessments of health equity. These inconsistencies undermine the reproducibility of research and limit the overall impact of proposed methodologies on diverse populations. Moreover, existing frameworks often neglect to assess the real-world health equity benefits of these models, a critical oversight given the potential to reinforce or mitigate healthcare disparities.

To address these challenges, we propose a two-fold approach: first, embedding fairness throughout the model development lifecycle to ensure that models are equitable from data collection to deployment; and second, expanding fairness reporting standards, such as TRIPOD-AI, to include explicit measures of bias detection and mitigation. With fairness considerations embedded into every stage of the MDLC—data collection, preprocessing, model training, evaluation, and deployment—we can ensure that biases are identified and addressed before they impact model performance or healthcare outcomes. Standardizing fairness reporting, especially with respect to health equity impacts, will further enhance transparency, improve reproducibility, and build trust with the diverse individuals whose data contribute to these models.

Moving forward, future research must prioritize closing these gaps by developing robust bias mitigation strategies and ensuring that models genuinely improve health equity. Adopting these recommendations allows us to create machine learning models that enhance clinical decision-making while also contributing to reducing healthcare disparities and advancing health equity for all.